\documentclass[a4paper]{article}
\pdfoutput=1
\usepackage{jhepmod}

\usepackage[utf8]{inputenc}

\usepackage[table ]{ xcolor}
\usepackage{enumitem}
\usepackage{multirow}
\usepackage{xcolor}
\usepackage{amsmath}
\usepackage{amsthm}

\usepackage{tikz}
\usetikzlibrary{shapes,arrows}
\tikzstyle{startstop} = [rectangle,rounded corners, minimum width=3cm,minimum height=1cm,text centered, draw=black,fill=red!30]
\tikzstyle{io} = [trapezium, trapezium left angle = 70,trapezium right angle=110,minimum width=3cm,minimum height=1cm,text centered,draw=black,fill=red!30]
\tikzstyle{process} = [rectangle,minimum width=3cm,minimum height=1cm,text centered,text width =3cm,draw=black,fill=orange!30]
\tikzstyle{decision} = [diamond,minimum width=3cm,minimum height=1cm,shape aspect=3,inner sep = 0.4pt,text centered,draw=black,fill=green!30]
\tikzstyle{arrow} = [thick,->,>=stealth]
\tikzstyle{shadow}=[preaction={fill=black,opacity=.5,transform canvas={xshift=0.5mm,yshift=-0.5mm},shading=radial,shading angle=20},fill=red]

\tikzstyle{ellipse}=[draw, rectangle, minimum width=2.8em, rounded corners=6pt,line width=0.5pt]% minimum height=1.5em, fill=red!20 椭圆
\tikzstyle{pxsbx}=[trapezium, trapezium left angle=75, trapezium right angle=105, minimum width=3em, text centered, draw = black, fill=white,line width=0.5pt] %平行四边形
\tikzstyle{lingxing}=[draw,diamond,shape aspect=3,inner sep = 0.4pt,thick,font=\itshape,line width=0.5pt]%,minimum size=8mm 菱形

\usepackage{amssymb}
\usepackage{graphicx,color}

\usepackage{amsmath,amsthm}

\usepackage{mathrsfs}
\usepackage{xcolor}

\usepackage{bm}

\def\beq{\begin{equation}}
\def\eeq{\end{equation}}
\newcommand{\bea}{\begin{eqnarray}}
\newcommand{\eea}{\end{eqnarray}}
\def\bi{\begin{itemize}}
\def\ei{\end{itemize}}
\def\ba{\begin{array}}
\def\ea{\end{array}}
\def\bfig{\begin{figure}}
\def\efig{\end{figure}}

\newcommand{\D}{\mathrm{d}}

\def\be{\begin{eqnarray}}
\def\ee{\end{eqnarray}}

\renewcommand{\o}{\omega}
\renewcommand{\O}{\Omega}

\newcommand{\lt}{\left}
\newcommand{\rt}{\right}

\title{Massive particles tunneling from quantum Oppenheimer-Snyder black holes and black hole entropy}

\author[1]{\ Hongwei Tan}  
\author[1]{\ Kui Xiao}  

\affiliation[1]{School of Science, Hunan Institute of Technology, Hengyang 421002, China}

\emailAdd{honweitan@hnit.edu.cn}
\emailAdd{Corresponding author:xiaokui@hnit.edu.cn}

\abstract{ 
    In this paper, we investigate the tunneling process of massive scalar particles from a quantum Oppenheimer-Snyder black hole using the tunneling approach proposed by Parikh and Wilczek. We compute the emission rate of these tunneling particles, which includes quantum correction terms compared to the result in classical General Relativity. These correction terms arise from loop quantum gravity effects. Following the scheme outlined in  \cite{zhang2008black,zhang2009black}, we derive the entropy of the black hole.
    Consistent with the universal black hole entropy formula in the context of quantum gravity, our findings include a logarithmic correction.
}

\begin{document}

\maketitle%\vspace{-7mm}

\section{Introduction}
Black hole (BH) is one crucial prediction of General Relativity (GR), which has attracted considerable attentions and undergone extensive study from both experimental and theoretical perspectives (see e.g., \cite{wald2010general, poisson2004relativist,frolov2011introduction,afshordi2024black,chandrasekhar1985mathematical}).
In the experimental realm, the gravitational wave detections \cite{LIGOScientific:2016aoc,LIGOScientific:2016sjg,LIGOScientific:2016dsl,LIGOScientific:2017bnn,abbott2016observation}  as well as the observations of the Event Horizon Telescope \cite{event2019first, akiyama2019first, akiyama2019first-2,akiyama2019first-3,akiyama2019first-4,akiyama2019first-5} provide compelling evidence for the existence of BHs.
In the future, more advanced instruments designed to detect BH signals, both in gravitational waves and electromagnetic waves, will be established, offering even stronger tests of the BH paradigm  \cite{Punturo:2010zz,Evans:2023euw,Evans:2021gyd,Colpi:2024xhw,TianQin:2015yph,Hu:2017mde}.
On the theoretical side, one of the most significant advances is BH thermodynamics, first introduced by \cite{bekenstein1973black}. 
In this work, the BH is treated as a thermodynamic system, with its temperature proportional to the surface gravity of the event horizon and its entropy given by
\begin{equation}\label{intro:BH_entropy}
    S=\frac{\mathscr{A}}{4l_p^2},\footnote{In this paper, the unit system is chosen as $C=G=1$.}
\end{equation}
with $l_p$ representing the Planck length and $\mathscr{A}$ the area  of the BH horizon.
This proposal is later refined by Hawking with quantum field theory in curved spacetime \cite{hawking1975particle}, where the Hawking radiation is predicted.
Despite the significant progress made in the study of BH physics, several unresolved issues remain, such as the black hole singularity problem and the information paradox (see e.g., \cite{penrose1965gravitational,hawking1970singularities,brady1995black,hawking1974black,hawking2005black, bekenstein2004black,landgren2019information}), where quantum gravity is expected to play a crucial role.

One competitive quantum gravity candidate is loop quantum gravity (LQG), a non-perturbative and background-independent
theory that may offer solutions to the difficulties mentioned previously \cite{thiemann2008modern, rovelli2004quantum,han2007fundamental,thiemann2003lectures,ashtekar2004background,giesel2012classical,rovelli2011zakopane,perez2013spin}.
Within the framework of LQG, the BH singularity is replaced by a big bounce, and thus resolves the BH singularity problem \cite{husain2022quantum,husain2022fate}.
Regarding BH entropy and the information paradox, the BH entropy is successfully reproduced by counting the number of the microstates of the BH horizon in the context of LQG \cite{rovelli1996black,perez2017black}.
Furthermore, the recently developed black-to-white hole transition approach is expected to resolve the BH information paradox \cite{rovelli2014planck, haggard2015quantum,de2016improved,christodoulou2016planck,bianchi2018white,PhysRevD.103.106014,soltani2021end,rignon2022black,han2023geometry}.

Since LQG is a candidate for quantum gravity, it is crucial to investigate its effective solutions, which are expected to provide phenomenological corrections to BH solutions within the framework of classical GR.
Recent developments in LQG corrections to the spherically symmetric self-gravitational collapse model have attracted remarkable attentions \cite{lewandowski2023quantum}.
This work is inspired by the pioneer of Oppenheimer and Snyder \cite{oppenheimer1939continued}, whose model, known as the Oppenheimer-Snyder (OS) model, was the first model to describe the self-gravitational collapse in classical GR.
In this model, the spacetime manifold is divided into an interior region and an exterior region.
The interior  region is filled with  pressure-less dust field, which satisfies the classical Friedmann equation.
The exterior region is vacuum, which is characterized by the
Schwarzschild solution. Due to its simplicity, the OS model can be solved exactly, providing profound insights on the nature of self-gravitational collapse.

The quantum version of the OS model is first proposed in \cite{lewandowski2023quantum}, which we refer to as the quantum OS (q-OS) model in this paper.
This model represents an intermediate stage of the self-gravitational collapse.
At the end of the collapse process, the collapsing phase transits to an expanding phase, leading to a black-to-white hole transition.
In this model, the interior region of the spacetime is described by the Ashtekar-Pawlowski-Singh (APS) model \cite{ashtekar2006quantum}, where the matter field satisfies the LQG-modified Friedmann equation.
To obtain the metric of the exterior region, the  Israel junction conditions are applied \cite{israel1966singular,shi2024higher}, yielding the following spacetime metric:
\begin{equation}\label{intro:qOS}
    \mathrm{d} s^2
    =-\left(1-\frac{2M}{r}+\frac{\alpha M^2}{r^4}\right)\mathrm{d}t^2
    +\left(1-\frac{2M}{r}+\frac{\alpha M^2}{r^4}\right)^{-1}\mathrm{d}r^2
    +r^2\mathrm{d}\O^2.
\end{equation}
Here $M$ is the ADM mass. $\alpha$ is a constant defined as $\alpha=16 \sqrt{3} \pi \gamma^3 l_p^2$, with $\gamma$ is the Barbero-Immirzi parameter-a dimensionless quantity.
The dimensional analysis indicates that $[\alpha]=[M]^2=[r]^2$.\footnote{$[A]$ denotes the dimension of the quantity $A$.}
Compared to the classical Schwarzschild solution, a correction $\frac{\alpha M^2}{r^4}$ appears in \eqref{intro:qOS}.
This term arises from LQG effects and may lead to significant and intriguing physical implications such as shadows and
quasinormal modes of the BH \cite{yang2023shadow,stashko2024quasinormal,zhang2023black,gong2024quasinormal,yang2024gravitational,liu2024gravitational,zi2024eccentric}.

One interesting implication of this LQG correction is the logarithmic correction to the BH entropy formula  \eqref{intro:BH_entropy}.
In our previous work \cite{tan2024black}, we examine the tunneling process of massless scalar particles from the q-OS BH with the Parikh-Wilczek (PW) approach \cite{parikh2000hawking, parikh2004secret}, which suggests that the Hawking radiation can be interpreted as a tunneling process.
In that work, we first compute the emission rate of the tunneling particles.
Then, by following the scheme in \cite{zhang2008black,zhang2009black}, we establish the BH entropy formula, which includes a logarithmic correction.
This result consists with the universal findings when the quantum gravity effects are taken into account (see e.g., \cite{ghosh2005log,kaul2000logarithmic,domagala2004black,meissner2004black,chatterjee2004universal,medved2004comment,lin2024effective, shi2024higher,banerjee2008quantum,banerjee2008quantum2,banerjee2009quantum,majhi2009fermion,majhi2010hawking}).

Naturally, this leads to further questions: What happens if we consider the tunneling process of massive particles? Does the logarithmic correction in the BH entropy formula still hold? In this paper, we address these questions. Specifically, we study the tunneling of massive process scalar particles from the event horizon of the q-OS black hole. Using the approach introduced in \cite{Zhang:2005sf}, we analyze the near-horizon behavior of massive particles, compute their emission rates, and ultimately derive the BH entropy formula within this framework. The resulting formula is as follows:
\begin{equation}
    S_{\text{OS}}=S_{\text{Sch}}+\frac{\pi\alpha}{2l_p^2}\log\left(\frac{\mathscr{A}_{\text{Sch}}}{l_p^2}\right)
    +O\left(\alpha^2\right),
    \end{equation}
    with $S_{\text{Sch}}$ is the entropy of the classical Schwarzschild BH.
    Our result also includes a logarithmic correction.
    This finding aligns with the universal behavior of black hole entropy formulas when quantum gravity effects are considered.

This paper is organized as following: 
In Sec. \ref{sec:rev_qOS}, we review briefly the q-OS model.
In Sec. \ref{sec:emission}, we compute the emission rate of the massive scalar particles tunneling from the event horizon of q-OS BH.
In Sec. \ref{sec:entropy}, we derive the entropy of the q-OS BH.
Finally, conclusions and future outlooks are presented in Sec. \ref{sec:om_out}.

\section{A brief review of the quantum Oppenheimer-Snyder black hole}\label{sec:rev_qOS}
The recently developed q-OS BH \cite{lewandowski2023quantum,yang2023shadow,stashko2024quasinormal,zhang2023black,gong2024quasinormal,yang2024gravitational,liu2024gravitational} is an effective model within the framework of LQG, inspired by the pioneer work of Oppenheimer and  Snyder \cite{oppenheimer1939continued}, which introduced the classical OS model.

The classical OS model  is the first model of self-gravitional collapse in GR.
In this model, the spacetime manifold $\mathcal{M}$ is split into an exterior region $\mathcal{M}^+$ and an interior region $\mathcal{M}^-$.
The exterior region is a spherical vacuum, described by the standard Schwarzschild metric.
The interior region is filled with the collapsing matter field and is described by the usual FRW metric
\begin{equation}
    \mathrm{d} s_{\mathrm{FRW}}^2=-\mathrm{d} \tau^2+a(\tau)^2\left(\mathrm{~d} \tilde{r}^2+\tilde{r}^2 \mathrm{~d} \Omega^2\right),
    \end{equation}
  where $\left(\tau,\,\tilde{r},\,\theta,\,\varphi\right)$ denotes a coordinate system.
  $a(\tau)$ is the scale factor, with which the Hubble constant $H:=\frac{\dot{a}}{a}$ can be defined.
The classical Friedmann equation then reads
  \begin{equation}\label{eq:class_fre_eq}
    H^2=\frac{8 \pi }{3} \rho,
    \end{equation}
    with $\rho$ is the density of the matter field.
  $a(\tau)$ satisfies the classical Friedmann equation.
To obtain the quantum version of the OS model, a LQG-deformed Friedmann equation is considered (see e.g., \cite{ashtekar2006quantum, yang2009alternative, assanioussi2018t})
\begin{equation}\label{eq:quan_fred}
    H^2=\frac{8 \pi }{3} \rho\left(1-\frac{\rho}{\rho_c}\right), 
    \end{equation}
$\rho_c$ is called the critical density, given by $\rho_c=\sqrt{3} /\left(32 \pi^2 \gamma^3\hbar\right)$.
Compared to the classical  Friedmann equation \eqref{eq:class_fre_eq}, there is a correction term $-\frac{8\pi\rho^2}{3\rho_c}$  in the LQG-deformed version.
By assuming the surface density of the matter field vanishes on $\Sigma$, with $\Sigma$ is a timelike hypersurface connecting $\mathcal{M}^+$ and $\mathcal{M}^-$, the Israel junction conditions give \cite{israel1966singular,shi2024higher}
\begin{equation}
    h_{\mu\nu}^+\left|_\Sigma\right.=h_{\mu\nu}^-\left|_{\Sigma}\right.,\quad
    K_{\mu\nu}^+\left|_\Sigma\right.=K_{\mu\nu}^-\left|_{\Sigma}\right..\footnote{The Greek letters $\mu,\,\nu$ are the spacetime indices, taking the value of $(0,\,1,\,2,\,3)$.   In the interior  region of spacetime, the coordinates are labeled as $x^0=\tau$, $x^1=\tilde{r}$, with $x^2$ and $x^3$ representing the angular directions. 
        In the exterior region, we use $x^0=t$, $x^1=r$, while $x^2$ and $x^3$ are the angular coordinates.}
\end{equation}
Here, the "$+$" sign denotes the exterior region while the "$-$" sign denotes the interior region.
$h_{\mu\nu}$ is the reduced metric on $\Sigma$, and $K_{\mu\nu}$ is the extrinsic curvature of $\Sigma$.
Then the metric of the exterior region of the q-OS spacetime reads
\begin{equation}\label{eq:quan_OS}
    \mathrm{d} s^2
    =-\left(1-\frac{2M}{r}+\frac{\alpha M^2}{r^4}\right)\mathrm{d}t^2
    +\left(1-\frac{2M}{r}+\frac{\alpha M^2}{r^4}\right)^{-1}\mathrm{d}r^2
    +r^2\mathrm{d}\O^2,
\end{equation}
with $M$ is the ADM mass of the matter field. Compared to the classical Schwarzschild solution, there is a correction term $\frac{\alpha M^2}{r^4}$ in \eqref{eq:quan_OS}, which is a quantum correction arises from LQG effects.
Eq. \eqref{eq:quan_OS} can be interpreted as an effective solution of LQG.

For a q-OS BH, the position of the BH's event horizon is determined by the solutions of 
\begin{equation}\label{eq:hori}
    f(r):=1-\frac{2M}{r}+\frac{\alpha M^2}{r^4}=0.
\end{equation}
As demonstrated in \cite{lewandowski2023quantum}, there are two real roots for eq. \eqref{eq:hori} when the ADM mass $M>M_{\text{min}}$, with $M_{\text{min}}$ is the lower bound for the mass required for real roots to exist. 
The lower bound is given by $M_{\min }:=\frac{4}{3 \sqrt{3} } \sqrt{\alpha}$.
In this paper, we focus on the process that massive scalar particles tunnel across the event horizon.
Note that $M\gg\alpha$, and by applying a Taylor expansion, the location of the event horizon is given by
\begin{equation}
    r_h=2M-\frac{\alpha}{8M}+O(\alpha^2)<2M.
\end{equation}
Detailed discussions on this issue can be found in \cite{tan2024black}.
\section{Massive particles tunneling from quantum Oppenheimer-Snyde black hole}\label{sec:emission}
In this section, we investigate the tunneling process of massive scalar particles from the  q-OS BH.
First, we rewrite the spacetime line element \eqref{eq:quan_OS} in Painlevé-Gullstrand coordinates \cite{painleve1921mecanique}.
Then, we examine the near horizon behaviors of massive particles.
Finally, we compute the emission rate of the outgoing massive scalar particles across the BH event horizon.
\subsection{Quantum Oppenheimer-Snyder black hole in Painlevé-Gullstrand coordinates}
To investigate the tunneling process, it is convenient to use a coordinate system that is regular at the BH horizons.
One of such coordinate systems is the Painlevé-Gullstrand coordinate system.
Firstly, we introduce the following coordinate transformations
\begin{equation}
    \lt(t,\,r,\,\theta,\,\varphi\rt)\to\lt(\tilde t,\,r,\,\theta,\,\varphi\rt).
\end{equation}
Here,  
\begin{equation}
    \tilde{t}=t+F(r),
\end{equation}
with 
\begin{equation}
    F(r)=\int_0^{r}\sqrt{\frac{M r^4\left(2 r'^3-M \alpha\right)}{\left(-2 M r'^3+r'^4+M^2 \alpha\right)^2}}\mathrm{d}r'.
\end{equation}
Then the line element \eqref{eq:quan_OS} is expressed in the Painlevé-Gullstrand coordinates as
\begin{equation}\label{eq:quan_OS_pain}
    \mathrm{d} s^2
    =-\left(1-\frac{2M}{r}+\frac{\alpha M^2}{r^4}\right)\mathrm{d}\tilde t^2
    +2\sqrt{\frac{2M}{r}-\frac{\alpha M^2}{r^4}}\mathrm{d}\tilde t\mathrm{d}r
    +\mathrm{d}r^2
    +r^2\mathrm{d}\O^2.
\end{equation}
A detailed derivation can be found in the appendix A of \cite{tan2024black}.
It is evident that \eqref{eq:quan_OS_pain} is regular at the BH event horizon.
Here we have chosen $+$ sign in front of $2\sqrt{\frac{2M}{r}-\frac{\alpha M^2}{r^4}}$, which describes the outgoing particles.
Furthermore, the timeslice of \eqref{eq:quan_OS_pain} is Euclidean,  allowing traditional Schr\"odinger equation remains valid.
Consequently, the WKB approximation also remains valid.
\subsection{Near horizon behaviors of massive particles}
To study the tunneling process of massive scalar particles,
it is crucial to examine their near horizon behavior, which differs significantly from that of massless particles. This difference arises because massive particles propogate along timelike geodesics.

To investigate the near horizon behaviors of massive particles, we follow the discussions in \cite{Zhang:2005sf}.
For simplicity, we assume the wave corresponding to the outgoing massive particles to be a de Broglie s-wave. 
Then, according to the WKB formula, the wave function reads
\begin{equation}\label{eq:wave_func}
    \Psi\left(r,t\right)
    =C\exp\left(i\left(\frac{1}{\hbar}\int_{r-\delta}^rp_r\mathrm{d} r-\o \tilde{t}\right)\right),
\end{equation}
where $r-\delta$ denotes the initial location of the particle.
$p_r$ here is the conjugate momentum of $r$.
Following \cite{Zhang:2005sf}, we let $\frac{1}{\hbar}\int_{r-\delta}^rp_r\mathrm{d} r-\o \tilde{t}=\phi_0$, with $\phi_0$ is a constant. 
By doing so, we fix the phase of the wave function as a constant in eq. \eqref{eq:wave_func}, which will not affect the physics under consideration. Next, we find
\begin{equation}
    \frac{\mathrm{d}r}{\mathrm{d}\tilde{t}}
    =\dot{r}
    =\frac{\o}{k},
\end{equation}
with $k$ is the de Broglie wave number, given by $k=\frac{p_r}{\hbar}$.
Therefore, $\dot{r}$ represents the phase velocity of the de Broglie wave, which can be expressed as
\begin{equation}\label{eq:dotr_tophase}
    \dot{r}=v_p.
\end{equation} 
Besides, the phase velocity $v_p$ and the group velocity $v_g$ satisfy the following relation:
\begin{equation}\label{eq:rel_gro_phs}
    v_p=\frac{1}{2}v_g.
\end{equation}
Hence, to compute $\dot{r}$, we need to find $v_g$.
As argued in \cite{Zhang:2005sf}, the events before and after the particle tunnels across the horizon can be considered as simultaneous. 
According to \cite{zheng2006synchronization}, for a general spacetime with a line element given by
\begin{equation}
    \D s^2=g_{00}\D t^2+2g_{0i}\D t \D x^i +g_{ij}\D x^i\D x^j \quad
    \left(i,\,j=1,\,2,\,3\right),
\end{equation}
the difference between the coordinates of time of two simultaneous events, occurring at different spatial points, is given by
\begin{equation}
    \D t=-\frac{g_{0i}}{g_{00}}\D x^i.
\end{equation}
In our case, the tunneling process of the particle can be equivalently viewed as the tunneling process of the corresponding wave package.
For convenience, we assume the particle tunnels radially, so $\D\theta=\D\varphi=0$.
Then with \eqref{eq:quan_OS_pain} we have 
\begin{equation}
    \D \tilde{t}=\frac{\sqrt{\frac{2M}{r}-\frac{\alpha M^2}{r^4}}}
    {1-\frac{2M}{r}+\frac{\alpha M^2}{r^4}}\D r_c,
\end{equation}
with $r_c$ denotes the position of the center of the wave package.
Then the group velocity  is given by 
\begin{equation}
    v_g:=\frac{\D r_c}{\D \tilde{t}}=\frac{1-\frac{2M}{r}+\frac{\alpha M^2}{r^4}}{\sqrt{\frac{2M}{r}-\frac{\alpha M^2}{r^4}}}.
\end{equation} 
Hence, with eqs. \eqref{eq:dotr_tophase} and \eqref{eq:rel_gro_phs}, we get 
\begin{equation}\label{eq:eom_for_par}
    \dot{r}=\frac{1-\frac{2M}{r}+\frac{\alpha M^2}{r^4}}{2\sqrt{\frac{2M}{r}-\frac{\alpha M^2}{r^4}}},
\end{equation} 
which is the equation of motion of outgoing massive particles near the horizon.
\subsection{Emission rate of massive particle tunneling}
PW approach suggests the Hawking radiation can be interpreted as a tunneling process across the BH horizons, where energy conservation and the back reaction of the Hawking radiation need to be taken into account \cite{parikh2000hawking}.
In this subsection, we consider the process of outgoing massive scalar particles tunneling across the event horizon of the q-OS BH.

As mentioned previously, we assume the de Broglie wave corresponding to the outgoing particle is a s-wave for simplicity.
In classical GR, Birkhoff's theorem states that any spherically symmetric solution of the vacuum field equations must be static and asymptotically flat.
In the q-OS model, there is a similar result by generalizing Birkhoff's theorem \cite{cafaro2024status}.
In their work, the authors demonstrate that the exterior region of the q-OS spacetime satisfies a  generalized Birkhoff’s theorem, with three possible cases corresponding to $k=-1,\,0,\,1$, representing open, flat and closed universes in the interior spacetime region respectively.
In our work, we focus on the case of $k=0$, as shown in \eqref{eq:class_fre_eq} and \eqref{eq:quan_fred}.
In this case, the exterior region of this type of q-OS model is unique and determined  by a single
parameter: the mass.
Thus, there is no gravitational wave emits in the tunneling process.
Suppose the energy of the tunneling particle is $\o$,
by energy conservation, the mass of the BH changes as $M\to M-\o$.
Therefore, line element \eqref{eq:quan_OS_pain} is modified as 
\begin{equation}\label{eq:quan_OS_pain_mod}
    \mathrm{d} s^2
    =-\left(1-\frac{2(M-\o)}{r}+\frac{\alpha (M-\o)^2}{r^4}\right)\mathrm{d}\tilde t^2
    +2\sqrt{\frac{2(M-\o)}{r}-\frac{\alpha (M-\o)^2}{r^4}}\mathrm{d}\tilde t\mathrm{d}r
    +\mathrm{d}r^2
    +r^2\mathrm{d}\O^2,
\end{equation}
as the particle is emitted.
Similarly, eq. \eqref{eq:eom_for_par} is modified as 
\begin{equation}\label{eq:modi_eom}
    \dot{r}=\frac{1-\frac{2(M-\o)}{r}+\frac{\alpha (M-\o)^2}{r^4}}
    {2\sqrt{\frac{2(M-\o)}{r}-\frac{\alpha (M-\o)^2}{r^4}}}.
\end{equation} 
Since the tunneling process occurs very close to the horizon, the outgoing particles experience an ever-increasing blue shift measured by local observers.
Accordingly, the geometric optics approximation can be applied,
and the WKB approximation is also valid.

By PW approach \cite{parikh2000hawking}, the outgoing particle is created inside the horizon,
then propagates across the horizon and reaches infinity.
While this is a classically forbidden process, it is allowed by the quantum tunneling approach. 
By WKB formula, the emission rate of the outgoing particle is given by 
\begin{equation}\label{eq:emi_rat}
    \Gamma \sim e^{-\frac{2}{\hbar} \operatorname{Im} A},
\end{equation}
with $A$ is the action of the particle.
Our focus is the imaginary part of $A$, computed as
\begin{equation}
    \operatorname{Im} A=\operatorname{Im} \int_{r_{\text {in }}}^{r_{\text {out }}} p_r \mathrm{d} r=\operatorname{Im} \int_{r_{\text {in }}}^{r_{\text {out }}} \int_0^{p_r} \mathrm{d}p_r^{\prime} \mathrm{d} r.
    \end{equation}
Here $r_{\text {in }}$ and $r_{\text {out }}$ represent the positions that before and after the particle tunnel across the horizon, respectively.
Hamiltonian equation gives
\begin{equation}
    \dot{r}=+\left.\frac{\mathrm{d} H}{\mathrm{d} p_r}\right|_r,
    \end{equation}
with the Hamiltonian $H=M-\o$.
Then with eq. \eqref{eq:modi_eom}, we obtain 
\begin{equation}\label{eq:imi_A}
    \begin{aligned}
        \operatorname{Im} A=&\operatorname{Im} \int_{r_{\text {in }}}^{r_{\text {out }}} \int_0^{p_r} \mathrm{d}p_r^{\prime} \mathrm{d} r\\
        =&-\operatorname{Im} \int_{r_{\text {in }}}^{r_{\text {out}}} \int_0^\omega \frac{\mathrm{d} \omega^{\prime}}{\dot{r}} \mathrm{~d} r\\
        =&-\operatorname{Im} \int_{r_{\text {in }}}^{r_{\text {out}}} \int_0^\omega \frac{2\sqrt{\frac{2(M-\omega)}{r}-\frac{\alpha (M-\omega)^2}{r^4}}}{1-\frac{2(M-\omega)}{r}+\frac{\alpha (M-\omega)^2}{r^4}}\mathrm{d} \omega^{\prime} \mathrm{~d} r.
    \end{aligned}
\end{equation}
Introduce $I=-\frac{2\sqrt{\frac{2(M-\omega)}{r}-\frac{\alpha (M-\omega)^2}{r^4}}}{1-\frac{2(M-\omega)}{r}+\frac{\alpha (M-\omega)^2}{r^4}}$, Taylor expansion implies
\begin{equation}\label{eq:taylor_I}
    I=-2\frac{\sqrt{\frac{2(M-\omega)}{r}}}{1-\frac{2(M-\omega)}{r}}
+\frac{\sqrt{\frac{M-\o}{r}}\lt(2(M-\o)^2+(M-\o)r\rt)}
{\sqrt{2}\lt(2(M-\o)-r\rt)^2r^2}\alpha
+O(\alpha^2).
\end{equation}
Note that $[\alpha]=[M]^2$, so $I$ is a dimensionless quantity. 
Therefore, every term on the right hand side of eq. \eqref{eq:taylor_I} is dimensionless.
In particular, it is easy to check that the quantity $\frac{\sqrt{\frac{M-\o}{r}}\lt(2(M-\o)^2+(M-\o)r\rt)}
{\sqrt{2}\lt(2(M-\o)-r\rt)^2r^2}\alpha$ is dimensionless. 
To compute \eqref{eq:imi_A}, we introduce a new variable $u=\sqrt{r}$.
Then 
\begin{equation}
    \begin{aligned}
        &\operatorname{Im} A\\
        =&-4\operatorname{Im} \int_{u_{\text {in }}}^{u_{\text {out }}} \int_0^\omega  \frac{\sqrt{2(M-\omega')}u^2}
        {(u+\sqrt{2(M-\omega')})(u-\sqrt{2(M-\omega')})}\mathrm{d} \omega^{\prime} \mathrm{~d} u\\
        &+2\operatorname{Im} \int_{u_{\text {in }}}^{u_{\text {out}}} \int_0^\omega
        \frac{(M-\omega')^{3/2}}{\sqrt{2}(2(M-\omega')-u^2)^2u^2}\alpha
         \mathrm{d}\omega^{\prime} \mathrm{d} u\\
         &+2\operatorname{Im} \int_{r_{\text {in }}}^{r_{\text {out}}} \int_0^\omega
         \frac{2(M-\omega')^{5/2}}
         {\sqrt{2}(2(M-\omega')-u^2)^2u^4}\alpha
         \mathrm{d} \omega^{\prime} \mathrm{d} u\\
         &+O\left(\alpha^2\right).
    \end{aligned}
\end{equation}
To avoid the singularity at $u=\sqrt{2(M-\omega')}$, we deform the countour of integration as $\o\to\o-i\epsilon$, with $\epsilon\to0$.
This ensures that he positive energy solutions decay over time \cite{parikh2000hawking}. As the consequence, we get
\begin{equation}\label{eq:int_with_deform}
    \begin{aligned}
        &\operatorname{Im} A\\
        =&-4\lim_{\epsilon\to0}\operatorname{Im} \int_{u_{\text {in }}}^{u_{\text {out }}} \int_0^\omega  \frac{\sqrt{2(M-\omega'+i\epsilon)}u^2}
        {(u+\sqrt{2(M-\omega'+i\epsilon)})(u-\sqrt{2(M-\omega'+i\epsilon)})}\mathrm{d} \omega^{\prime} \mathrm{~d} u\\
        &+2\lim_{\epsilon\to0}\operatorname{Im} \int_{u_{\text {in }}}^{u_{\text {out}}} \int_0^\omega
        \frac{(M-\omega'+i\epsilon)^{3/2}}{\sqrt{2}(2(M-\omega'+i\epsilon)-u^2)^2u^2}\alpha
         \mathrm{d}\omega^{\prime} \mathrm{d} u\\
         &+2\lim_{\epsilon\to0}\operatorname{Im} \int_{r_{\text {in }}}^{r_{\text {out}}} \int_0^\omega
         \frac{2(M-\omega'+i\epsilon)^{5/2}}
         {\sqrt{2}(2(M-\omega'+i\epsilon)-u^2)^2u^4}\alpha
         \mathrm{d} \omega^{\prime} \mathrm{d} u\\
         &+O\left(\alpha^2\right).
    \end{aligned}
\end{equation}
We observe that eq. \eqref{eq:int_with_deform} {converges uniformly.
Hence, the integration and summation commutes.
Therefore,
\begin{equation}
    \begin{aligned}
        &\operatorname{Im} A\\
        =&4\pi\int_0^\omega (M-\omega')\mathrm{d} \omega^{\prime}
        +\frac{3\pi}{16}\lim_{\epsilon\to0}\operatorname{Im}
        \int_0^\omega\frac{i}
          {(M-\omega'+i\epsilon)}
      \alpha
       \mathrm{d}\omega^{\prime} \\
       +&\frac{5\pi}{16}\lim_{\epsilon\to0}\operatorname{Im}  \int_0^\omega
       \frac{i}
       {M-\omega'+i\epsilon}\alpha
   \mathrm{d} \omega^{\prime}
   +O\left(\alpha^2\right).
    \end{aligned}
\end{equation}    
Note that we are considering the BH mass  with the scale of Solar mass, so $M\gg\o$.
Therefore,
\begin{equation}
    \begin{aligned}
        \operatorname{Im} A
        =&4\pi M\lt(\omega-\frac{\omega^2}{2M}\rt)
        +\frac{\pi}{2}\alpha\lt(\log\lt(\frac{M}{l_p}\rt)-\log\lt(\frac{M-\omega}{l_p}\rt)\rt)
        +O\left(\alpha^2\right)
        \\
        =&4\pi M\lt(\omega-\frac{\omega^2}{2M}\rt)
        +\frac{\pi}{4}\alpha\lt(\log\lt(\frac{\mathscr{A}_{\text{Sch}}(M)}{l_p^2}\rt)-\log\lt(\frac{\mathscr{A}_{\text{Sch}}(M-\omega)}{l_p^2}\rt)\rt)
        +O\left(\alpha^2\right),
    \end{aligned}
\end{equation}
where $\mathscr{A}_{\text{Sch}}(M)$ is the area of the classical Schwarzschild BH with mass $M$. 
Compared to the typical result in \cite{parikh2000hawking}, there are logarithmic corrections appear in our finding. 
These corrections arise from the LQG effects.
Then with eq. \eqref{eq:emi_rat}, we find the emission rate of the outgoing particle as 
\begin{equation}\label{eq:emmi_rate_res}
\Gamma \sim\exp
\left[
    -\frac{8\pi M}{l_p^2}
    \lt(\omega-\frac{\omega^2}{2M}\rt)
        -\frac{\pi\alpha}{2l_p^2}\lt(\log\lt(\frac{\mathscr{A}_{\text{Sch}}(M)}{l_p^2}\rt)-\log\lt(\frac{\mathscr{A}_{\text{Sch}}(M-\omega)}{l_p^2}\rt)\rt)
        +O\left(\alpha^2\right)
\right].
\end{equation}
\section{Comments on the black hole entropy} \label{sec:entropy}
As demonstrated in \cite{zhang2009black,zhang2008black}, if the emission rate of the particles can be expressed as
\begin{equation}\label{eq:emiss_in_entro}
    \Gamma\sim\exp{\Delta S},
\end{equation}   
 the tunneling process implies the BH evaporation process is unitary.
Here $\Delta S$ is the change of the BH entropy before and after the tunneling process occours. 
Concret examples indicate that the emmison rates in classical GR satisfy \eqref{eq:emiss_in_entro}, see for example \cite{parikh2000hawking,Zhang:2005sf, Zhang:2005wn}.
To express the emission rate \eqref{eq:emmi_rate_res} in the formalism of \eqref{eq:emiss_in_entro}, we introdoce entropy of the q-OS BH as 
\begin{equation}\label{eq:OS_entropy}
S_{\text{OS}}=S_{\text{Sch}}+\frac{\pi\alpha}{2l_p^2}\log\left(\frac{\mathscr{A}_{\text{Sch}}}{l_p^2}\right)+O\left(\alpha^2\right),
\end{equation}
with $S_{\text{Sch}}$ is the entropy of the classical Schwarzschild BH.
Finally, the emission
rate reads
\begin{equation}
    \Gamma\sim\exp{\Delta S_{\text{OS}}}.
\end{equation}   
There is a logarithmic correction appears in our result \eqref{eq:OS_entropy}, consisting with the universal results when quantum gravity effects are taken into account
 \cite{ghosh2005log,kaul2000logarithmic,domagala2004black,meissner2004black,chatterjee2004universal,medved2004comment,lin2024effective, shi2024higher,banerjee2008quantum,banerjee2008quantum2,banerjee2009quantum,majhi2009fermion,majhi2010hawking,nozari2008hawking, amelino2006black, nozari2006comparison,nozari2012natural,ashtekar2003quantum,corichi2007polymer}.
In particular, it consists with the result within in our previous work \cite{tan2024black}, where we consider the massless scalar particles tunnel from the q-OS BH.
 The sign of the prefactor for the logarithmic correction in \eqref{eq:OS_entropy} is positive, which differs from the results in the framework of ordinary LQG \cite{ghosh2005log,kaul2000logarithmic,domagala2004black}.
However, Our finding is consistent with the BH entropy formula of the effective loop quantum BH obtained by other approaches \cite{lin2024effective, shi2024higher}. 
Indeed, the prefactor of the logarithmic correction in the BH entropy formula is deeply related to the micro-states of the BH horizon.
Ref. \cite{medved2004comment} provides profound insight into this issue.
On the one hand, the logarithmic corrections arising from quantum gravity effects will reduce the BH entropy.
On the other hand, the logarithmic corrections riginating from thermal fluctuations will increase the BH entropy, due to the thermal fluctuations enhance the uncertainty of the micro-states of the BH horizon.
Accordingly, our findings might suggest that the logarithmic correction in the entropy formula of the q-OS BH arises from both quantum gravity effects and (effective) thermal fluctuations.
Suppose $a_q$ is the logarithmic prefactor corresponding to the quantum gravity effect.
This is a negative constant as argued in \cite{medved2004comment}. 
Hence, to obtain the total prefactor $+\frac{\pi\alpha}{2l_p^2}$ in  \eqref{eq:OS_entropy}, the logarithmic prefactor corresponding to the (effective) thermal flutuations must be present, which is denoted as $a_t$ in this paper.
Then the total logarithmic prefactor is given by $a_q+a_t=+\frac{\pi\alpha}{2l_p^2}$.
We will continue to explore these interesting topics in our future researches.
\section{Conclusions and outlooks}\label{sec:om_out}
In this paper, we investigate the tunneling process of massive scalar particles from the  q-OS BH.
We  apply the PW approach incoorporate with the near horizon behaviors of massive particles proposed in \cite{Zhang:2005sf}.
Compared to the results in the framework of classical GR \cite{parikh2000hawking,parikh2004secret}, our findings reveal quantum corrections, which arise from the constribution of LQG.
following the methodology outlined in \cite{zhang2009black,zhang2008black}, we establish the entropy of the q-OS BH, which includes a logarithmic correction to the entropy of the classical Schwarzschild BH.
Our findings consist with the universal results when quantum gravity effect is considered 
\cite{ghosh2005log,kaul2000logarithmic,domagala2004black,meissner2004black,chatterjee2004universal,medved2004comment,lin2024effective, shi2024higher,banerjee2008quantum,banerjee2008quantum2,banerjee2009quantum,majhi2009fermion,majhi2010hawking,nozari2008hawking, amelino2006black, nozari2006comparison,nozari2012natural,ashtekar2003quantum,corichi2007polymer}.

In particular, the sign of the prefactor of the logarithmic correction is consistent with the results for loop quantum BH obtained by other approaches \cite{lin2024effective, shi2024higher,tan2024black}. 
The sign of the prefactor of the logarithmic correction is a crucial issue in quantum gravity.
As demonstrated by previously works \cite{ghosh2005log,kaul2000logarithmic,domagala2004black}, the sign of the prefactor is negative in the framework of LQG, aligning with the arguments in \cite{medved2004comment}.
However, the sign in both of our finding in present paper and in the context of the effective loop quantum BH approach ( as presented in\cite{lin2024effective, shi2024higher}) is positive.
This suggests that the positive prefactor may represent a universal result.

Investigating this issue could provide new insights into quantum gravity theory and the black hole information paradox. One possible interpretation is that the logarithmic correction to the entropy formula in the context of effective loop quantum BHs includes contributions from both quantum gravity effects and (effective) thermal fluctuations.
Specifically, we denote the prefactor of the logarithmic correction arising from quantum gravity effects as $a_{q}$ and the prefactor of the logarithmic correction arising from (effective) thermal fluctuations as $a_t$.
The previous findings indicate $a_q+a_t>0$ in the framework of loop quantum BH.
We plan to explore this challenging but important topic in future research, as it may yield a deeper understanding of the microstates of the BH horizon.

Besides the tunneling approach, there are many other manners to compute the BH entropy. 
Such as the brick wall model \cite{hooft1985quantum,hooft1996scattering} and the Hamiltonian-Jacobi approach \cite{bergamin2008black}, which have been well established in previous works.
In particular, the recently developed island scheme has garnered significant attention, as it may potentially resolve the black hole information paradox \cite{penington2020entanglement,almheiri2019entropy,almheiri2020page,almheiri2019islands,hashimoto2020islands}.
In our future research, we aim to apply these approaches to the q-OS BH in order to compute its entropy and compare the results to those obtained via the PW approach. These studies could provide profound insights into the theory of quantum gravity.
\section{Acknowledgment}
HT acknowledges the valuable discussions with Xiangdong Zhang and Rong-zhen Guo.
The authors acknowledge the support from the Hunan Provincial Natural Science Foundation of China (Grant No. 2022JJ30220).
\bibliographystyle{jhep}
\bibliography{paper}
\end{document}